\documentclass[webpdf,modern,large,namedate,final]{oup-authoring-template}
\usepackage{float}
\hypersetup{colorlinks, citecolor={blue}, urlcolor={blue}}

\graphicspath{{Fig/}}

\theoremstyle{thmstyleone}%
\theoremstyle{thmstyletwo}%
\theoremstyle{thmstylethree}%

\begin{document}

\journaltitle{Bioinformatics}
\DOI{https://doi.org/10.1093/bioinformatics/XXXXXXX}
\copyrightyear{2023}
\pubyear{2023}
\access{Advance Access Publication Date: Day Month Year}
\appnotes{Application Note}

\firstpage{1}

\title[GOLEM]{GOLEM: distribution of Gene regulatOry eLEMents within the plant promoters}

\author[1]{Lukáš Nevosád}
\author[2]{Božena Klodová\ORCID{0000-0003-4229-2868}}
\author[2]{David Honys\ORCID{0000-0002-6848-4887}}
\author[3,4]{Radka Svobodová\ORCID{0000-0002-3840-8760}}
\author[3,4]{Tomáš Raček\ORCID{0000-0002-0296-2452}}
\author[3,4,$\ast$]{Petra Procházková Schrumpfová\ORCID{0000-0003-0066-1581}}

\authormark{Lukáš Nevosád et al.}

\address[1]{\orgname{Tripomatic s.r.o.}, \orgaddress{\street{Za Parkem 14}, \postcode{60200}, \state{Brno}, \country{Czech Republic}}}
\address[2]{\orgname{Institute of Experimental Botany of the Czech Academy of Sciences}, \orgaddress{\street{Rozvojová 263}, \postcode{165 02}, \state{Prague}, \country{Czech Republic}}}
\address[3]{\orgdiv{National Centre for Biomolecular Research}, \orgname{Faculty of Science, Masaryk University}, \orgaddress{\street{Kamenice 5}, \postcode{625 00}, \state{Brno}, \country{Czech Republic}}}
\address[4]{\orgdiv{CEITEC -- Central European Institute of Technology}, \orgname{Masaryk University}, \orgaddress{\street{Kamenice 5}, \postcode{625 00}, \state{Brno}, \country{Czech Republic}}}

\corresp[$\ast$]{Corresponding author. \href{email:schpetra@sci.muni.cz}{schpetra@sci.muni.cz}}

%\received{Date}{0}{Year}
%\revised{Date}{0}{Year}
%\accepted{Date}{0}{Year}

%\editor{Associate Editor: Name}

\abstract{\textbf{Motivation:} The regulation of gene expression during tissue development is extremely complex. One of the key regulatory mechanisms of gene expression involves the recognition of regulatory motifs by various proteins in the promoter regions of many genes. Localisation of these motifs in proximity to the transcription start site (TSS) or translation start site (ATG) is critical for regulating the initiation and rate of transcription. The levels of transcription of individual genes, regulated by these motifs, can vary significantly in different tissues and developmental stages, especially during tightly regulated processes such as sexual reproduction. However, the precise localisation and visualisation of the regulatory motifs within gene promoters with respect to gene transcription in specific tissues, can be challenging.\\
\textbf{Results:} Here, we introduce a program called GOLEM (Gene regulatOry eLEMents) which enables users to precisely locate any motif of interest with respect to TSS or ATG within the relevant plant genomes across the plant Tree of Life (\textit{Marchantia}, \textit{Physcomitrium}, \textit{Amborella}, \textit{Oryza}, \textit{Zea}, \textit{Solanum} and \textit{Arabidopsis}). The visualisation of the motifs is performed with respect to the transcript levels of particular genes in leaves and male reproductive tissues, and can be compared with genome-wide distribution regardless of the transcription level.\\ 
\textbf{Availability and implementation:} GOLEM is freely available at \url{https://golem.ncbr.muni.cz} and its source codes are provided under the MIT licence at GitHub at \url{https://github.com/sb-ncbr/golem}.
}

\maketitle

\section{Introduction}
The regulation of gene expression is a dynamic process that requires a tightly orchestrated control mechanism. The precise timing and magnitude of gene transcription, especially during plant development, can be regulated by several DNA sequence motifs. These regulatory motifs can be found in the vicinity of the transcription start sites (TSS) - upstream or even downstream, including the five-prime untranslated region (5’ UTR) - and serve as binding sites for various proteins. These proteins act as molecular switches that can activate or repress gene expression \citep{galli2020mapping, schmitz2022cis}.

Dynamic changes in gene transcription are essential, especially during the development of plant reproductive tissues. Several short DNA motifs involved in the regulation of key genes required for the differentiation of male germline, active in sperm cells and pollen vegetative cells, have already been identified: LAT52 pollen-specific motif in tomato (AGAAA) \citep{bate1998functional}; LAT52/59 core motif (GTGA)~\citep{twell1991promoter}; DOF core motif (AAAG)  \citep{yanagisawa2002dof, li2014characterization}; ARR1AT motif (NGATT) \citep{sharma2011putative}; pollen Q-element (AGGTCA) \citep{hamilton1998monocot}, and many others \citep{sharma2011putative,peters2017conserved, hoffmann2017cis, li2014characterization}. Furthermore, even common motifs like the TATA box (TATAWA), which are involved in initiating transcription, exhibit multiple variants that may contribute to the establishment of unique expression patterns \citep{bernard2010tc}.

Transcriptomics has gained significant popularity in recent years, as it has the ability to provide detailed insights into gene expression dynamics across different tissues, developmental stages, or experimental conditions \citep{tyagi2022upcoming}. Transcriptome sequencing (RNA-Seq) is an important method for investigating gene regulation. While most transcriptome analyses have traditionally focused on easily accessible plant materials such as leaves or seedlings, there is a growing trend towards exploring transcriptomes from intricate and deeply embedded tissues, including sperm cells and various pollen developmental stages \citep{julca2021comparative, klodova2022regulatory}.

Achieving precise localisation and visualisation of regulatory motifs in genes that exhibit high transcription levels in specific tissues requires a multi-step procedure, which may be limited by the user’s familiarity with various bioinformatics software. We present a user-friendly online software GOLEM (Gene regulatOry eLEMents), which allows browsing various tissues such as sporophyte (leaves) or male gametophyte developmental tissues (antheridia, pollen stages, sperm cells) across the relevant plant genomes within the plant Tree of Life. Our software enables us to investigate the precise localisation and distribution of motifs of interest in gene promoters, which can be specified by the level of gene expression in specific tissues. Furthermore, tracking of the genome-wide distribution across mosses, monocots and dicots, regardless of the transcription level, may aid to track the evolution of regulatory motifs. Additionally, the set of genes with specific motifs at defined positions can be exported for further analysis. 

\section{Materials and methods/Algorithm}

The GOLEM program is divided into two main phases: data processing pipeline and data visualisation.  

\subsection{Data processing}

The reference genomes and annotations files from \textit{Marchantia polymorpha}, \textit{Physcomitrium patens}, \textit{Amborella trichopoda}, \textit{Oryza sativa}, \textit{Zea mays}, \textit{Solanum lycopersicum}, and \textit{Arabidopsis thaliana} were downloaded in the FASTA format and general feature (GFF3) format, respectively (Supplementary Data 1). The data processing pipeline first parses genomic location data of single genes provided in GFF3 files. The GFF3 file is used to identify the position of transcription initiation (five\_prime\_UTR) and translation initiation (start\_codon), i.e. determines the positions of TSS (transcription start site) and ATG (first translated codon) in the reference genome. A fixed number of base pairs before and after the TSS or ATG are included in the dataset. 

Next, the pipeline matches selected genes against a Transcript Per Million (TPM) table from various tissues and developmental stages. The TPM values express normalized transcription rates of each gene obtained from RNA-seq data sets. The TPM values for \textit{M. polymorpha}, \textit{P. patens}, \textit{A. trichopoda}, \textit{O. sativa}, and \textit{S. lycopersicum} tissue were acquired from Conekt database (\url{https://conekt.sbs.ntu.edu.sg}) \citep{julca2021comparative}. The TPM values for pollen developmental stages of \textit{A. thaliana} Columbia (Col-0) and Landsberg erecta (Ler) were extracted from \citep{klodova2022regulatory}. The TPM values for \textit{Z. mays} and some of the \textit{A. thaliana} developmental stages (e.g. leaves, seedlings, tapetum) were established from the FASTQ files using the pipeline described in Supplementary Data 2. The normalised TPM values used for the data processing pipeline are listed in a table in Supplementary Data 3.

\begin{figure}[htbp!]
\includegraphics[width=\linewidth]{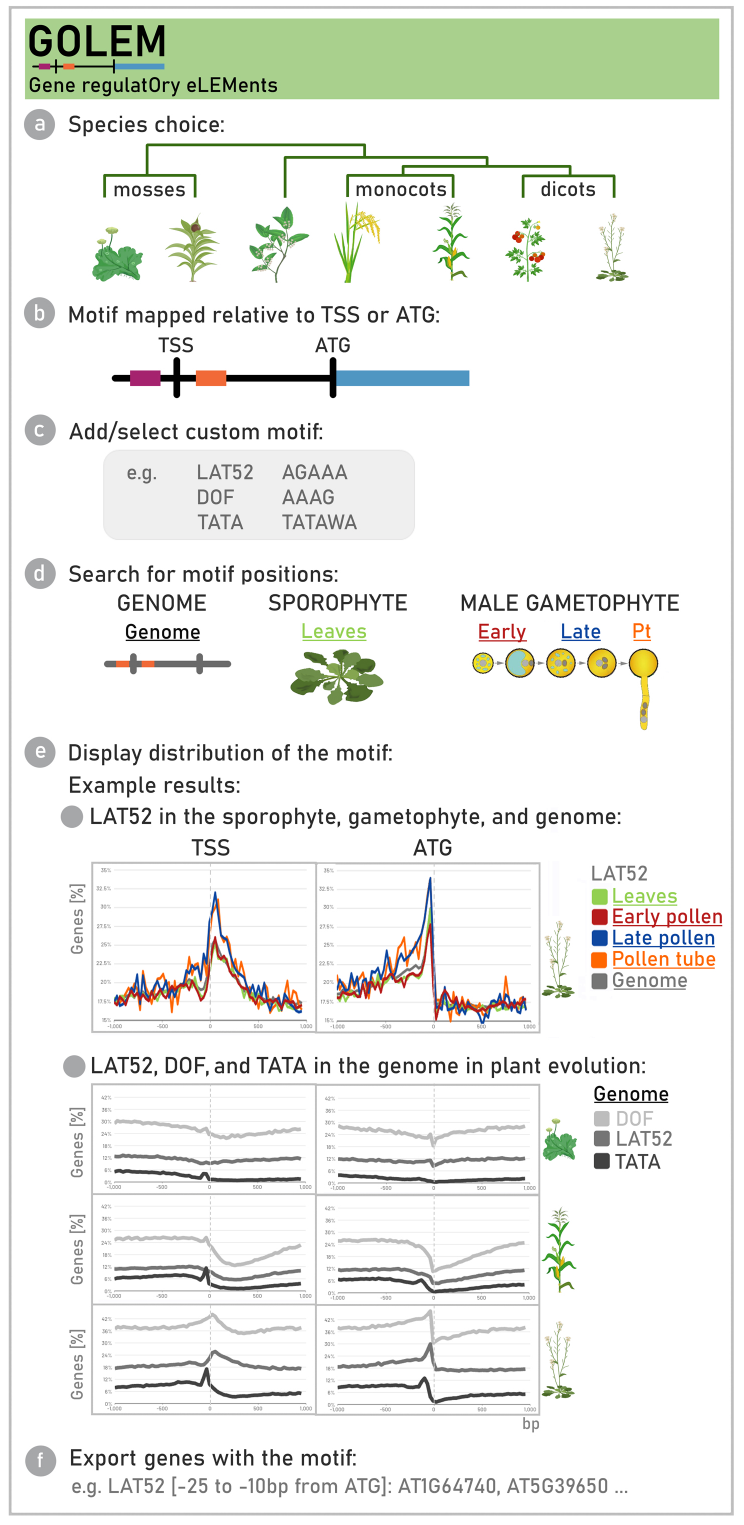}
\caption{An illustrative workflow of the GOLEM software. (a) Genome data for chosen species is downloaded on the web browser (\textit{Marchantia}, \textit{Physcomitrium}, \textit{Amborella}, \textit{Oryza}, \textit{Zea}, \textit{Solanum}, and \textit{Arabidopsis}). (b) Region in the vicinity of TSS or ATG is specified. (c) Motifs of interest are defined. (d) Leaves or male tissues and the transcriptional levels of the genes included in the analysis are defined, with the option to analyse genome-wide distribution regardless of transcription. (e) The exemplified LAT52 motif shows distribution downstream of TSS but upstream of ATG, with higher prevalence in the promoters of genes transcribed during Late pollen development. The genome-wide distribution of DOF, LAT52, and TATA-box motifs enables their tracking across the evolutionary tree. (f) The accession numbers of genes with certain motif within a defined region and tissue may be exported in XLSX format tables.}
\label{fig:workflow}
\end{figure}

The output of the data processing pipeline is a separate FASTA-compatible file that contains each valid gene from the original input, along with information about the position of TSS and ATG, and transcription rates in each stage added as comments. The pipeline also generates a validation log that provides information about genes that were excluded, i.e. non-protein coding genes (noStartCodonFound), pseudogenes (noFivePrimeUtrFound, noTpmDataFound), genes without TSS (noFivePrimeUtrFound, if relevant for certain organism).

Motif search uses regular expressions to search the input string of base pairs. For each motif, the reverse complement was calculated and then translated both into regular expressions. When the regular expression is run against the input data, we record all results and calculate their relative positions (adjusted to the middle of the motif) relative to TSS and ATG. 

\subsection{Data visualisation}

The data visualisation phase consists of five steps, as depicted in Fig.~\ref{fig:workflow}. In the selected genome (Fig.~\ref{fig:workflow}a), the user can then choose the genomic interval to be searched, effectively specifying the window of the sequence where the search is performed, and focus on a defined region in the vicinity of TSS or ATG (Fig.~\ref{fig:workflow}b). The user can also select motifs for the search, including custom motifs and preset motifs resulting from previous research. Motifs are searched in both forward and reverse forms, and the reverse form is calculated automatically (Fig.~\ref{fig:workflow}c).

Before the entire analysis, the user confirms the tissue and developmental stages to be searched and selects the method for selecting genes for analysis (Fig.~\ref{fig:workflow}d). Gene selection uses a given percentile (default is the 90th percentile) to select the genes that are the most/least transcribed in tissues or developmental stages of interest to exclude the genes with low or even negligible transcription. The number of selected genes within the given percentile can be tracked during the proceeding steps. Besides the given percentile, the number of genes included in the analysis may be determined by the user. Regardless of the transcription level, the motif distribution can also be included in the analysis (stage Genome).

The goal of the analysis is to visualise the distribution of motifs of interest in the vicinity of the TSS or ATG of all protein-coding genes, or exclusively in selected genes that exhibit high/low transcription levels in particular tissues and developmental stages (Fig.~\ref{fig:workflow}e). Results are presented graphically, with each stage colour-coded, and can be displayed as simple counts or percentages. The user can also choose to display either the number of motifs found or aggregate the motifs by genes.

Additionally, the set of genes with specific motifs at defined positions can be exported for further analysis (Fig.~\ref{fig:workflow}f). Within the analysis, the application allows the user to export each data series or export the aggregated data for all data series in XLSX format. The user can also see the distribution of individual motifs and drill down through them.

The data visualisation phase involves an application written in Flutter/Dart \citep{meiller2021modern}, which can be run as a standalone application or compiled into JavaScript and hosted on the web as a single-page web application (\url{https://golem.ncbr.muni.cz}).

\subsection{Limitations}

The entire processing takes place on the client within the application or web browser, with input files loaded into memory. The program’s ability to work with large datasets may be constrained by the available memory and the web browser's local client settings. We found that even in the web application, where performance is limited due to the inefficiencies of JavaScript compared to platform native code, performance is satisfactory on modern computers without the need for significant code optimisation.

\section{Results}
Gene expression is primarily controlled through the specific binding of various proteins to diverse DNA sequence motifs. Our software GOLEM is a convenient tool for visualising the distribution of motifs of interest in the vicinity of TSS or ATG and identifying genes expressed at a certain level within selected tissues.

The distributions of one core promoter motif TATA (TATA-box) \citep{bernard2010tc}, together with two pollen-associated motifs -- LAT52 (LAT52 pollen-specific motif from tomato, POLLEN1LeLAT52) \citep{bate1998functional} and DOF core motif (DOFCORE) \citep{yanagisawa2002dof, li2014characterization} -- are exemplified in Fig.~\ref{fig:workflow}e. Our analysis revealed that the preferential position of the LAT52 motif (the top of the peak) is located downstream of the TSS and upstream to ATG, i.e. in the five-prime untranslated region (5’ UTR). Moreover, the number of the genes containing the LAT52 in \textit{A. thaliana} is higher within the 5’ UTR region of genes exhibiting elevated expression (90th percentile) in Late pollen development and in Pollen tubes, as opposed to genes expressed during Early pollen development or in Leaves.

Additionally, our analysis has shown that the TATA-box resides upstream of the TSS in all three exemplified genomes across the evolutionary tree (moss \textit{P. patens}, monocot \textit{Z. mays} and dicot \textit{A. thaliana}). On the other hand, the pollen-associated motifs DOF and LAT52 show in \textit{A. thaliana} a sharp peak centred downstream from the TSS.

From the genes with LAT52 motif expressed in the Late pollen stage of \textit{A. thaliana}, the XLSX format table was exported. From this table two pollen-associated genes (ATG164740, AT5G39650), with LAT52 motif between -25 to -10 base pair (bp) area, are exemplified in Fig. 1f. Generally, the accession numbers of the genes exported from GOLEM can be analysed by various bioinformatical approaches: e.g. Gene description search; Gene ontology (GO) enrichment analysis; Differential expression analysis; Protein-protein interaction network analysis.

\section{Conclusion}

To address the challenge of precise localisation and visualisation of regulatory motifs near transcription and translation start sites, which play a critical role in gene regulation in specific tissues, we introduce the GOLEM software. Our software offers a user-friendly platform to investigate the distribution of any motif of interest in gene promoters across diverse plant genomes and developmental tissues.

\section{Acknowledgement}

Biological Data Management and Analysis Core Facility of CEITEC Masaryk University, funded by ELIXIR CZ research infrastructure (MEYS Grant No: LM2023055), is gratefully acknowledged for supporting the research presented in this paper. 

\section{Author contribution}
P.P.S. and D.H. conceived the study. B.K. analysed RNA-seq data and calculated the TPM. L.N. imposed computation analysis and visualisation tool. T.R. and R.S. implemented the website and supported program accessibility. P.P.S. wrote the paper with the help of all co-authors. 

\section{Declaration of Competing Interest}
The authors report no declarations of interest.

\section{Funding}
This work was supported for P.P.S. and D.H. by the Czech Science Foundation [21-15841S] and for B.K. by the Czech Science Foundation [21-15856S]; and collaboration with Tripomatic s.r.o. was supported by the European Regional Development Fund Project ‘SINGING PLANT’ [CZ.02.01/0.0/0.0/16\_026/0008446].  

\bibliographystyle{abbrvnat}
\bibliography{golem}

\end{document}